\documentclass[aps,prc,preprint,floatfix,showpacs,amsmath,amssymb,superscriptaddress,nofootinbib]{revtex4-1}

\usepackage[utf8]{inputenc}
\usepackage{braket, amsmath}
\usepackage{mathtools}
\usepackage{tikz}
\usetikzlibrary{quantikz}
\usepackage{adjustbox}
\usepackage{color}
\usepackage[]{xcolor}
\usepackage{graphicx}
\usepackage{siunitx}
\usepackage{appendix}
\usepackage{dcolumn}% Align table columns on decimal point
\usepackage{bm}% bold math
\usepackage{array}
\usepackage[bookmarks=true]{hyperref}
\usepackage{multirow}

\usepackage{braket}
\usepackage{caption}
\usepackage{subcaption}
\usepackage{academicons}
\usepackage{multirow}
\usepackage{upgreek}
\newcommand{\orcid}[1]{\href{https://orcid.org/#1}{\includegraphics[width=10pt]{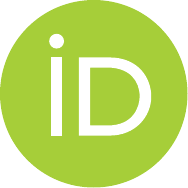}}}
\hypersetup{
	colorlinks=true,
	pdfborder={0 0 0},
	citecolor=blue,
	linkcolor=blue,
	urlcolor=blue
}

\begin{document}
\title{A Quantum Algorithm for the Linear Response of Nuclei}

\author{Abhishek\orcid{0000-0003-2226-3146}}
\email{abhishek@ph.iitr.ac.in}
\affiliation{%
 Department of Physics, Indian Institute of Technology Roorkee, Roorkee
247 667, India 
}%

\author{Nifeeya Singh}
\email{n\_singh@ph.iitr.ac.in}
\affiliation{%
 Department of Physics, Indian Institute of Technology Roorkee, Roorkee
247 667, India 
}%

\author{Pooja Siwach\orcid{0000-0001-6186-0555}}
\email{psiwach@physics.wisc.edu}
\affiliation{%
 Department of Physics, University of Wisconsin-Madison, Madison, Wisconsin 53706, USA 
}%

\author{P. Arumugam\orcid{0000-0001-9624-8024}}%
 \email{arumugam@ph.iitr.ac.in}
%  \orcid{skjdbfks}
\affiliation{%
 Department of Physics, Indian Institute of Technology Roorkee, Roorkee
247 667, India 
}%

\date{\today}
%%%%%%%%%%%%%%%%%%%%%%%%%%%%%%%%%%%%%%%%%

\begin{abstract}
We present a quantum algorithm to obtain the response of the atomic nucleus to a small external electromagnetic perturbation. The Hamiltonian of the system is presented by a harmonic oscillator, and the linear combination of unitaries (LCU) based method is utilized to simulate the Hamiltonian on the quantum computer. The output of the Hamiltonian simulation is utilized in calculating the dipole response with the SWAP test algorithm. The results of the response function computed using the quantum algorithm are compared with the experimental data and provide a good agreement. We show the results for $^{120}$Sn and $^{208}$Pb to corroborate with the experimental data in Sn and Pb region and also compare the results with those obtained using the conventional linear response theory.
\end{abstract}

\maketitle

\section{INTRODUCTION}\label{intro}
With the advent of NISQ (near-term intermediate scale quantum) era quantum computers, several successful attempts have been made to perform the quantum simulations of nuclear many-body problems~\cite{Dumitrescu:2018, Roggero:2019, Roggero:2020, Denis:2020, Pooja:qc, cervia2021,Pooja:2022, Andres:2022, Romero:2022, Hobday:2022,Denis:2022}.
The simulations of dynamics of quantum many-body systems, especially the response of a many-body system to an external perturbation, on the classical (conventional) computers have well-known computational limitations due to exponential growth in the Hilbert space. Quantum computers are a promising tool to scale these simulations for a large enough system. One of the most useful applications of such an approach is the giant dipole resonances (GDRs)~\cite{Pring,Harakeh,rhine_2015,arumugam2004} in the atomic nuclei. GDR is the hallmark of collective motion inside a nucleus. GDR represents an effective probe into the response of a nucleus subjected to a weak electromagnetic external perturbation. Quantum mechanically, it is the transition between the ground state and a collective excited state.

Since its discovery in 1937~\cite{first_detect}, there has been an extensive study in the field of GDR, both experimentally~\cite{exp1berman,CARLOS197461,DONALDSON2018133,rhine_exp2013,rhine_exp2016} and theoretically~\cite{theory1982,rhine_2015,arumugam2004,arumugam2005,theory2005,theory2011,theory2021,QRPA1,RRPA1}. Random phase approximation (RPA)~\cite{DJRowe} is the most widely used theoretical model for such collective excitation in nuclei. Finding the RPA solutions depends on the matrix formulation of the RPA equations, where several complex residual interactions are involved. The problem is simplified by considering a mean-field approximation that assumes an independent particle picture~\cite{Pring,maruhn} and a separable residual interaction between the nucleons. Quantum computers, on the other hand, deal with the quantum many-body states directly, which can prove a better probe in the more fundamental exploration of collective behavior shown by every quantum many-body system. Hence developing such algorithms becomes crucial, and atomic nuclei provide the best tool to test them as there is an extensive amount of experimental data available for the GDR~\cite{DIETRICH1988199,exp1berman} for almost all the available nuclei.

In this work, we propose an approach to calculate the nuclear response for the GDR excitations on a quantum computer. Our approach consists of two parts, the first part includes the simulation of the Hamiltonian of the system using the linear combinations of unitaries (LCU)~\cite{Childs2012HamiltonianSU,LCU2} method and the second part is to utilize the output of the Hamiltonian simulation (energies and wave functions) to compute the dipole response function on the quantum computer using the SWAP test~\cite{ripper2022swap} by calculating the overlap between the ground and excited states. We prepare the excited states using a method proposed in Ref.~\cite{excited_roggero}. We utilize a dipole operator as the excitation operator and calculate the probability of excitation using the SWAP test. These probabilities are then used in calculating the final response of the nucleus. We compare our results with the already established theoretical models in the classical computation regime such as the linear response theory (LRT)~\cite{Pring,Ring_1983,Ring_1984,gallardo} which is based on the random phase approximation (RPA)~\cite{maruhn,DJRowe}. LRT successfully reproduces the GDR built on the ground state of the nucleus, and it uses the single-particle wave functions and energies to get a superposition of possible particle-hole (ph) excitations which gives the final collective excitation known as the giant resonance. On the quantum computer, the states are no longer single-particle states instead, they represent the quantum many-body states of a second quantized Hamiltonian~\cite{neilsen_chuang}. The ground state in such representation is also known as the quantum vacuum state, and all other states represent the excited states of the system. Hence, in qubit space, the response function is taken as the superposition of all the possible excitations from the vacuum state to the excited states, similar to Ref.~\cite{chemistry_response}. 

We compare the results from both the LRT-based approach computed on the classical computers and the quantum computation approach with the available experimental data. We compute the results for $^{120}$Sn and $^{208}$Pb to benchmark our approach on the nuclear chart for Sn and Pb region. We utilize the harmonic oscillator (H.O.) potential to represent the mean field in classical computation and its second quantized form to calculate the response on a quantum computer. Our method is currently limited to the spherical nuclei as a deformed system requires a larger basis size which increases the number of required qubits in our approach to unfeasible levels for the currently available hardware. 

The paper is arranged as follows. In Sec.~\ref{theoretical}, we detail our theoretical framework for the quantum algorithm and the LRT. In Sec.~\ref{RnD}, we discuss our results and then conclude in Sec.~\ref{conclude}.

\section{Theoretical Framework}\label{theoretical}
We calculate the response of the nucleus under the external perturbation, where the microscopic degrees of freedom for the nucleus are calculated using quantum algorithms. Here we give our formalism by first discussing the Hamiltonian of our system and its implementation on a quantum computer, followed by the description of the response function. Finally, we briefly describe quantum algorithms utilized in the present work to obtain the response function on a quantum computer.

\subsection{Hamiltonian}
The Hamiltonian of the system in the second quantization notation  can be written as 
\begin{equation}\label{Hamiltonian}
    H = \sum_{ij} h_{ij} a^{\dagger}_i a_j ,
\end{equation}
where $h_{ij}$ are the matrix elements of $H$ in a single-particle basis
\begin{equation}\label{hij}
   h_{ij} = \bra{i} (\hat{T}+\hat{V}) \ket{j}.
\end{equation}
$\ket{i}$ represents the basis of the many-body system such that a single-particle eigenstate can be expanded as $\ket{\psi_{sp}} = \sum_{i}c_i\ket{i}$. For a given form of Hamiltonian, $h_{ij}$ are calculated with classical computation, and $a^{\dagger}_i, a_{i}$ are evaluated on a quantum computer after transforming them into quantum gates using quantum transformations like Jordan-Wigner (JW) transformation. The JW transformation is a mapping of quantum many-body basis in Fock-space to the qubit's Hilbert space and can be defined as~\cite{JW:1928}
\begin{align}\label{JW_encoding}
    a^{\dagger}_{j} &= \frac{1}{2}(X_j-\iota Y_j) \otimes \prod_{k<j}Z_k\;,  \\
    a_{j} &= \frac{1}{2}(X_j+\iota Y_j) \otimes \prod_{k<j}Z_k\;,
\end{align}
where $X, Y,$ and $Z$ represent the Pauli gates. 
In this work, we utilize the harmonic oscillator (H.O.) Hamiltonian which is diagonalized in the H.O. basis, such that $h_{ij}$ is diagonal in the single-particle basis. i.e.,
\begin{equation}
   h_{N'N} = \bra{N'} (\hat{T}+\hat{V}) \ket{N} = \hbar\omega\left[\left(N+\frac{3}{2}\right)\delta_{N}^{N'} \right] 
\end{equation}
Here $\ket{N}$ represents the H.O. eigen basis where $N$ determines the size of the basis. Truncating the basis size to N = 4 and 5, for Jordan-Wigner transformation, the associated Hamiltonians $H_4$ and $H_5$ using Eq.~(\ref{Hamiltonian}) can be written as
\begin{equation}
    H_4 = (6I-0.75Z_0-1.25Z_1-1.75Z_2-2.25Z_3)\hbar\omega
\end{equation}
\begin{equation}
    H_5 = (8.75I-0.75Z_0-1.25Z_1-1.75Z_2-2.25Z_3-2.75Z_4)\hbar\omega
\end{equation}
where $\hbar\omega = 41 A^{\frac{-1}{3}}$~\cite{Pring}. The formulas for $H_{N}$ corresponding to greater values of $N$ can be found easily by following the JW transformation. As we increase the basis size to diagonalize the Hamiltonian, the number of qubits required to represent it in the qubit space also increases. 

\subsection{Nuclear Response Function}
The response of the nucleus under external perturbation is calculated using the linear response theory (LRT) based on the framework of random phase approximation (RPA). More details about RPA and LRT can be found in Refs.~\cite{gallardo,Ring_1983,maruhn}.  We use a separable interaction of the form 
\begin{equation}\label{hhat}
    \hat{H^{'}}= \hat{H}+\frac{1}{2}\sum_{\alpha=1}^3~\kappa_{\alpha}\hat{D}^{\dag}_{\alpha}\hat{D}_{\alpha},
\end{equation}
where $\hat{H}$ is the single-particle Hamiltonian [Eq.~(\ref{Hamiltonian})]. $\hat{D}_{\alpha}$ is the single-particle dipole operator where $\alpha$ represents the three spatial directions and $\kappa_{\alpha}$ is the strength parameter of the dipole-dipole interaction. $\hat{D}_{\alpha}$ is defined as~\cite{Ring_1983}
\begin{equation}\label{d_alpha}
    D_{\alpha}=\frac{NZ}{A}(r_{com}^N-r_{com}^P),
\end{equation}
where $r_{com}^N$ and $r_{com}^P$ are the centers of mass of neutrons and protons, respectively. Here, $N$, $Z$, and $A$ are the neutron, proton, and atomic mass numbers, respectively. The response function matrix $R$ can be calculated using the linearized Bethe-Salpeter equation~\cite{Pring,gallardo,Ring_1984}, with matrix elements given by 
\begin{equation}\label{Rij}
    R_{\alpha\beta}=\frac{R^0_{\alpha\beta}}{1-R^0_{\alpha\alpha}\kappa_{\alpha}}.
\end{equation}
The $R^0$, the response function without the residual interaction, is given by~\cite{gallardo}
\begin{align}\label{R0}
    R^{0}_{\alpha\beta}&=\sum_{pq}\frac{\bra{q}\hat{D}_{\alpha}\ket{p}\bra{p}\hat{D}^{\dag}_{\beta}\ket{q}}{E-(\epsilon_q-\epsilon_p)+i \Gamma}(n_q-n_p).
\end{align}
$\ket{q}$ and $\ket{p}$ are single-particle states with energies $ \epsilon_q$ and $\epsilon_p $  and occupation numbers $n_q$ and $n_p$, respectively. In the case of classical computation, the response function utilizes the single-particle wave functions and energies obtained from the mean field. On the other hand, when we map the problem on the qubits, the states are no longer the single-particle states but lie in Fock space~\cite{neilsen_chuang} as many-body states. Hence to perform the quantum computation, the $R^{0}$ is also calculated with these many-body states similar to Ref.~\cite{chemistry_response} as
\begin{equation}\label{R0_quantum}
    R^{0}_{\alpha\beta} = \sum_{\nu} \frac{\bra{0}\hat{D}_{\alpha}\ket{\Psi_{\nu}}\bra{\Psi_{\nu}}\hat{D}^{\dag}_{\beta}\ket{0}}{E-(e_{\nu}-e_0)+i \Gamma},
\end{equation}
where $e_{\nu}$ is the energy of $\nu^{th}$ excited state in the qubit space with $\nu = 0$ representing the ground or vacuum state. $\ket{\Psi_{\nu}}$ represents the corresponding many-body wave vector for the $\nu^{th}$ excited state. 
The value of the parameter $\Gamma$ is chosen as 2.0 MeV to reproduce the experimental width. Self consistent values of coupling strengths $\kappa_{\alpha}$ are~\cite{gallardo} 
\begin{align}\label{kappa}
    \kappa_{\alpha}=\kappa \frac{3A}{NZ}M\omega^2(\alpha),
\end{align}
where $\omega^2(\alpha)$ are the oscillator frequencies corresponding to the structure of the nucleus calculated with the $\hat{H_0}$, and are given by $\hbar(\omega(1)\omega(2)\omega(3))^{\frac{1}{3}}=41A^{\frac{-1}{3}}$~\cite{gallardo}. Here $\omega^{'}s$ are inversely proportional to the semi-axes lengths $R(1) = R(\theta = \frac{\pi}{2}, \phi = 0), R(2) = R(\theta = \frac{\pi}{2}, \phi = \frac{\pi}{2}), R(3) = R(\theta = 0, \phi = \frac{\pi}{2})$, where $R(\theta, \phi)$ defined as
\begin{equation}
R(\theta,\phi)=\mathcal{C}R_0\left[\sum_{\lambda=0}^{\lambda_{max}}\sum_{\mu=-\lambda}^{\lambda} a_{\lambda\mu}Y_{\lambda\mu}(\theta,\phi)
\right]~.
\label{shape}
\end{equation}
Here $R_0$ is  the radius of the equivalent spherical nucleus with the same
volume, $a_{\lambda\mu}$ are deformation parameters, $Y_{\lambda\mu}$ are the spherical harmonics, and $\mathcal{C}$ is  the volume conservation constant. $M$ is the mass of the nucleon. 

The cross-section in the intrinsic frame is related to the response function as~\cite{gallardo}

\begin{align}\label{sigma}
    \sigma(E)&= \frac{-4\pi e^2 E}{\hbar c}\sum_{\alpha=1}^3 \text{Imag}(R_{{\alpha}{\alpha}}),
\end{align}
where $E$ is interpreted as the incident energy in the photo-absorption cross-section, and \textit{c} is the speed of light in the vacuum. For a non-rotating system, the cross-section in intrinsic and lab frames are equal.

\subsection {State preparation}\label{state_prepare}
\begin{figure}
    \centering
    \includegraphics[width = \textwidth]{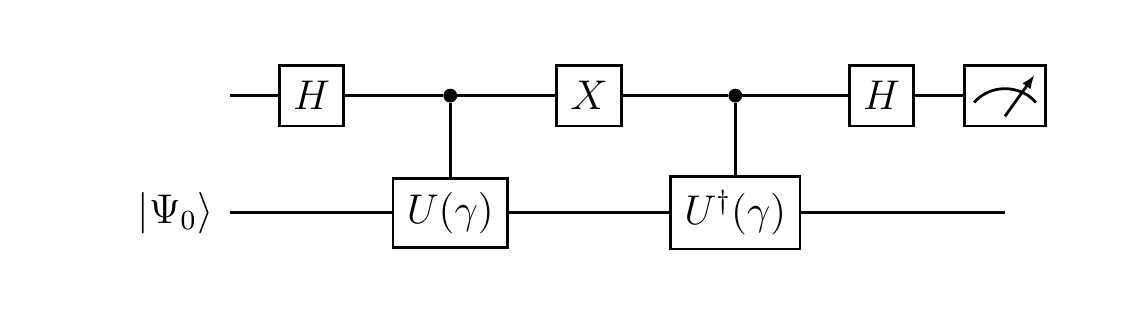}
    \caption{Quantum circuit for state preparation.}
    \label{fig:excited_state_circuit}
\end{figure}

We outline the state preparation technique in this section. We begin by initializing a quantum register in the desired quantum state {$\ket{\Psi_0}$}. In Ref.~\cite{Roggero:2020}, the algorithm from the time-dependent state preparation method was put forth. The primary goal is to produce an approximate state using the time-evolution operator linked to the excitation operator $O$.
\begin{equation}
    U(\gamma) = \exp(-i\gamma O) = \cos{\gamma O}-i\sin{\gamma O}.
\end{equation}
Such that the approximate state $\ket{\Psi_A(\gamma)}$ is given by
\begin{equation}
    {\ket{\Psi_A(\gamma)}}\propto \sin(\gamma O) = \ket{\phi_E}+\mathcal{O}(\gamma^2).
\end{equation}
The ``time'' argument in this case, is $\gamma$. We carry out this operation with the circuit depicted in Fig.~\ref{fig:excited_state_circuit} using the unitary $U(\gamma)$, which is managed by an ancilla qubit.
The final state created by preparing the ancilla qubit in the state $\ket{0}$ and using the circuit is
\begin{equation}\label{excited}
    \ket{\Omega(\gamma)} = \ket{0}\otimes \cos{(\gamma O)}{\ket{\Psi_0}} - i\ket{1}\otimes \sin{(\gamma O)}{\ket{\Psi_0}}
\end{equation}
Now, we can choose only the second component of the state in Eq.~(\ref{excited}) by using the spatial measurement on the ancilla qubit shown in the circuit in Fig.~\ref{fig:excited_state_circuit}. As a result, the system register will remain in the state Eq.~(\ref{excitedd}) if the measurement identifies the ancilla as being in the state $\ket{1}$.
\begin{equation}\label{excitedd}
    {\ket{\Psi_A(\gamma)}} = \frac{-i}{\sqrt{{\bra{\Psi_0}}\sin^2(\gamma O){\ket{\Psi_0}}}}\sin(\gamma O){\ket{\Psi_0}}
\end{equation}
If we measure the ancilla in $\ket{0}$, we start the process over from scratch and make another attempt.
In this technique, the excitation operator for a single qubit is a dipole operator. The reaction's initial state is defined as ${\ket{\Psi_0}}$, and its end state is $\ket{1}$. A minor state preparation inaccuracy necessitates a shorter period($\gamma$). And small values of $\gamma$ will, however, also have a small success probability. Therefore, we must confine the parameter. More specifications are provided in Ref.~\cite{Roggero:2020}. As an alternative to this time-dependent strategy, we can also prepare the excited state using a method based on a linear combination of unitaries~\cite{Roggero:2020,Childs2012HamiltonianSU}.  

\subsection{Calculation of Overlaps}
\begin{figure}
    \centering
    \includegraphics[width = 10.6cm]{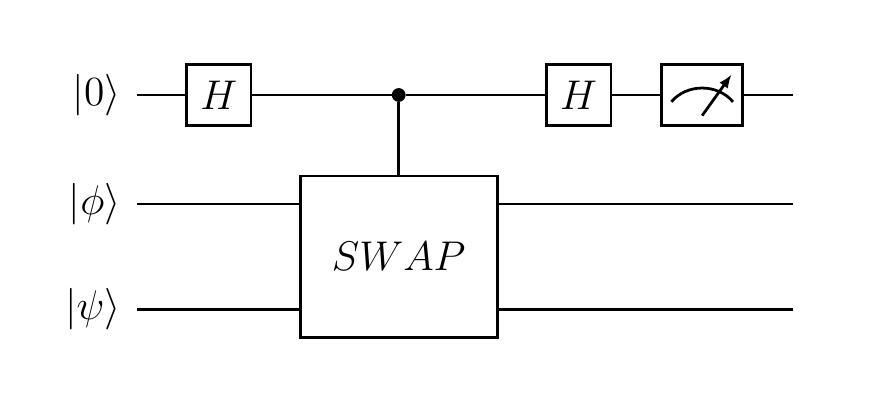}
    \caption{Quantum circuit that performs the SWAP test.}
    \label{fig:swap_circuit}
\end{figure}

\begin{figure}
    \centering
    \includegraphics[width = \textwidth]{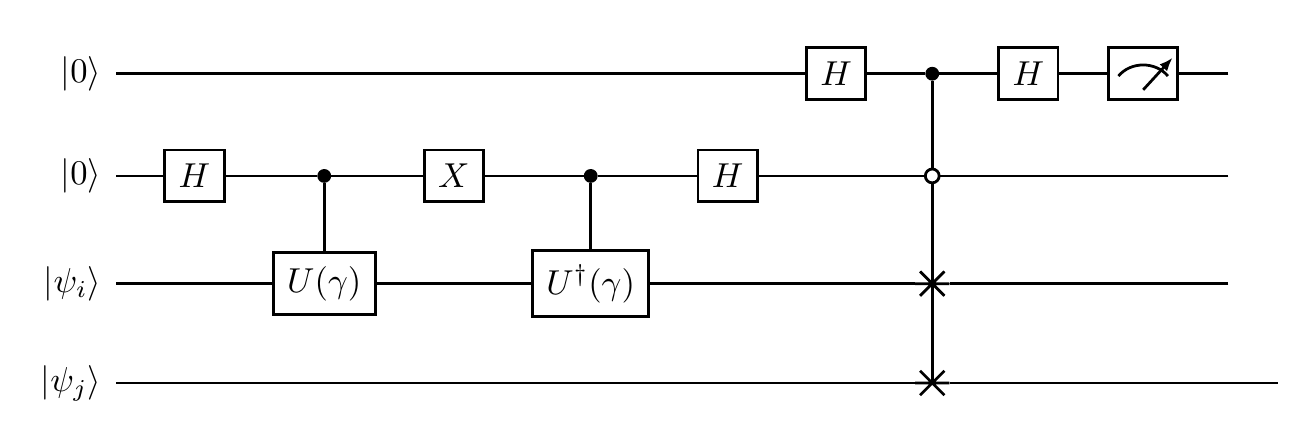}
    \caption{Complete quantum circuit}
    \label{fig:overlap_circuit}
\end{figure}

To evaluate the overlap, we implemented the swap test. We can establish some amount of certainty about how two states are comparable using the SWAP~\cite{2013garcia,ripper2022swap}. Fig.~\ref{fig:swap_circuit} depicts the quantum circuit utilized in the test. Two equal-dimension states $\ket{\psi}$ and  $\ket{\phi}$ and an auxiliary qubit in the $\ket{0}$ state constitute the inputs, and three gates (two Hadamard gates (H), one controlled SWAP gate (CSWAP)) are used. Here $\ket{\phi}$ represents the operator-operated excited state obtained in the section~\ref{state_prepare} and the overall circuit is shown in Fig.~\ref{fig:overlap_circuit}.
The state of the auxiliary qubit is evaluated at the circuit's end. When the $\ket{0}$ state is obtained, this situation is referred to as outcome 0, and when the $\ket{1}$ state is obtained, this situation is referred to as outcome 1. The outcome is 0 with a probability 1 if the states are equal, i.e., $\ket{\phi} = \ket{\psi}$. The difference between the probabilities of outcome 0 and outcome 1 yields the overlap ${\lvert \braket{\phi|\psi} \rvert}^2$. 

\subsection{Calculation of Energies}
\begin{figure}
    \centering
    \includegraphics[width = 10.6cm]{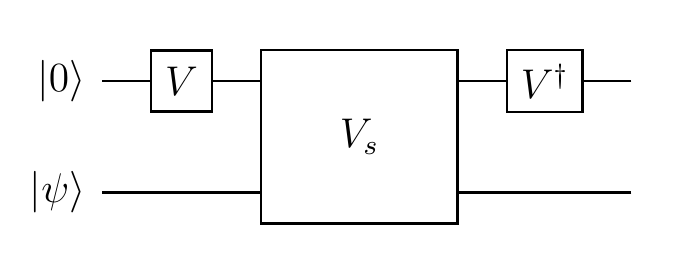}
    \caption{Quantum circuit to calculate $H\ket{\psi}$ with the method based on the linear combination of unitaries.}
    \label{fig:v_circuit}
\end{figure}

Numerous techniques, including the linear combination of unitaries (LCU)~\cite{Childs2012HamiltonianSU}, variational quantum eigensolver  (VQE)~\cite{VQE_nature1,VQE_nature2} algorithm, and quantum phase estimation (QPE)~\cite{neilsen_chuang}, can be used to calculate the energy values. Here, we adopt the LCU approach because the QPE calls for more ancilla qubits for more precise findings, leading to increasing circuit depth and running time. On the other hand, VQE is used mostly for finding ground-state energy. To calculate the non-unitary operation on the wave function, we use the LCU algorithm. This algorithm can also be utilized to prepare an excited state for a nuclear system (an alternative to Sec.~\ref{state_prepare}). Given a Hamiltonian as linear combination of unitaries $H = \sum_{i=0}^{k-1} a_i U_j$ with $a_i > 0$ $\forall i \in [0,k-1]$.
We first apply $V$ on an ancilla qubit. Then, using the ancilla as the control, we perform a controlled $U$ gate $(V_{s})$ on the state $\ket{i}$. Finally, we measure the ancilla qubit in the computational basis by applying $V^{\dagger}$ to it as given in Ref.~\cite{Pooja:2022}. The Quantum circuit corresponding to it is shown in Fig.~\ref{fig:v_circuit}.

The first step is to prepare a state which requires that we define operator $V$ as

\begin{equation}
    V\ket{0}^{\otimes n_a} = \frac{1}{\sqrt{\Lambda}}\sum_{i}\sqrt{\ a_i}\ket{i},
\end{equation}
where $\Lambda =\sum_{i} {a_i}$. For example, let us consider a Hamiltonian $ H = a_0 U_0+a_1 U_1$ then 
\begin{equation}
    V\ket{0} = \frac{\sqrt{a_0}}{\sqrt{a_0 + a_1}}\ket{0} + \frac{\sqrt{a_1}}{\sqrt{a_0 + a_1}}\ket{1}.
\end{equation}
The next step is to apply $ V_s =  \sum_{i=0}^{L}\ket{i}\bra{i}\otimes U_i$ operator also known as select operator~\cite{Pooja:2022}, which is applied on both the ancilla and the state qubits. 
Next is to apply $V^{\dagger}$ and measure the ancilla qubit in $\ket{0}$ state. This will give the energy of $\ket{\psi}$ state.
By computing the overlap of $\ket{\psi}$ and $H\ket{\psi}$ with the aid of the swap test, as previously described, we can also find the expected value $\braket{\psi|H|\psi}$. By taking the eigenstates of our Hamiltonian as $\ket{\psi}$, we can calculate the energy spectrum of our system.

\section{RESULTS AND DISCUSSION}\label{RnD}
\begin{figure}
    \centering
    \includegraphics[width = 0.6\textwidth]{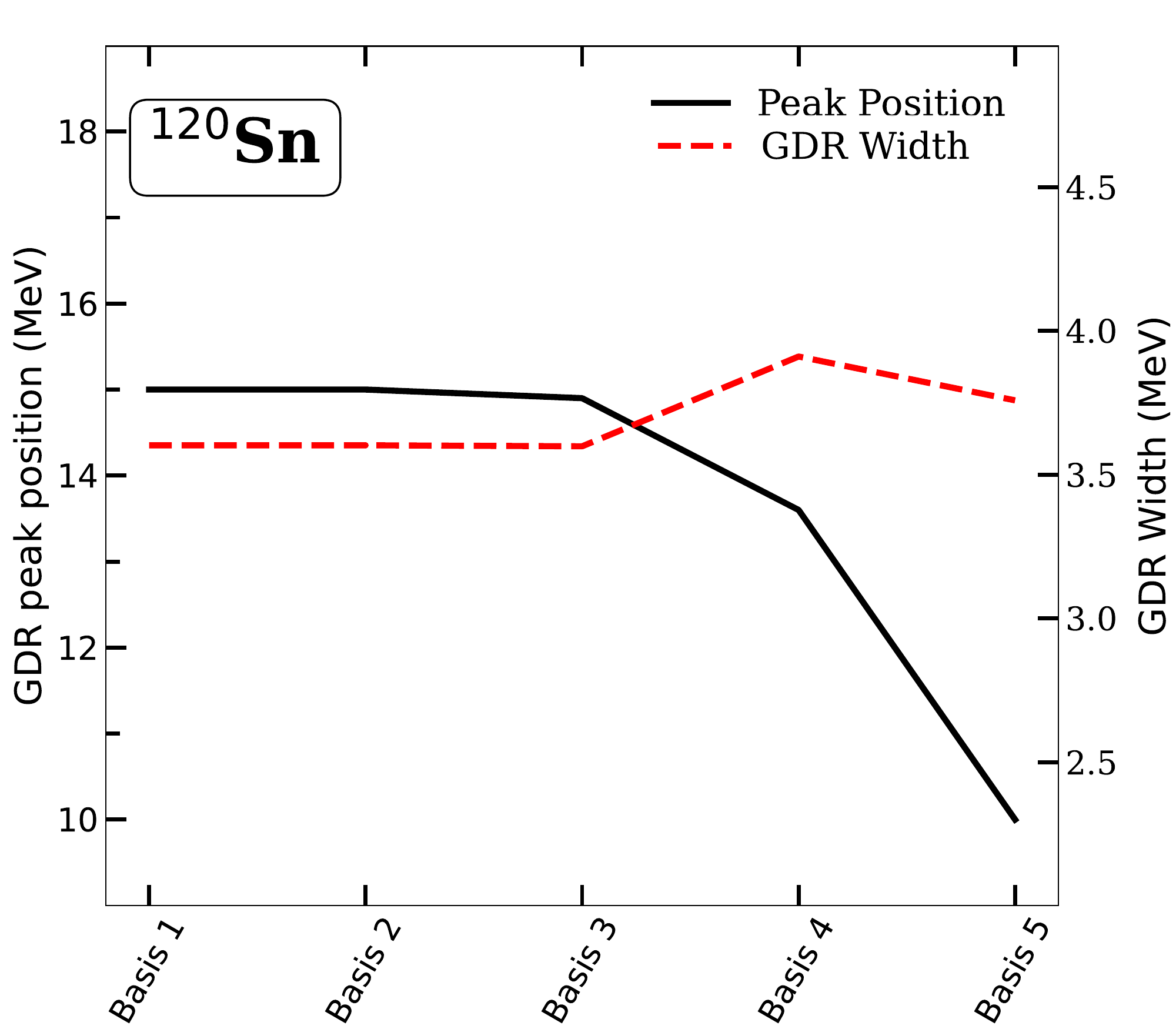}
    \caption{For $^{120}$Sn, the variation of GDR peak and width with the choice of basis size is obtained using classical computation.}
    \label{fig:peak_width_sn}
\end{figure}

We discuss our results for two nuclei, the $^{120}$Sn and $^{208}$Pb, to show the versatility of our method across the nuclear chart. In the case of classical computation, the results are obtained within the H.O. basis with a basis size where the value of the principle quantum number $N$ is taken from 0 to 10. However, from the results of classical computation, we know that the states which are near the Fermi level dominate the response function of the nucleus, as shown in Fig.~\ref{fig:peak_width_sn}. The different labels on the X-axis are related to the different choices of basis sizes, as shown in Table~\ref{basis}. The change in the position of the GDR peak and in the value of GDR width remains insignificant up to basis 3 ($N$ = 3 to 6). Further reduction after basis 3 leads to a significant shift in the GDR peak position and a jump in the GDR width. Hence in our calculations, instead of using the full basis, we use basis 3 to obtain the response function on the Quantum computer for the case of $^{120}$Sn. The Fermi level for $^{120}$Sn is represented by the $N = 4$ state as every state has a degeneracy of $(N+1)(N+2)$ and thus only states above and below the Fermi level dominate the GDR response function. Similar to the case of $^{120}$Sn, we also utilize the states near the Fermi level for the case of $^{208}$Pb where the proton and neutron Fermi levels are given by $N = 5$ and $N = 6$ states, respectively. The benefit of using a smaller basis ( 4 H.O. states) is that we only need four qubits to represent the quantum state of our system in the qubit space. The value of dipole-dipole strength parameter $\kappa$ is chosen as 0.4 for the results of the classical computation in both the nuclei and for the results of quantum computation, it is taken as 0.5 and 0.85 for $^{120}$Sn and $^{208}$Pb, respectively. Such a reduction in dipole-dipole strength ($\kappa < 1$) is necessary to reproduce the experimental results similar to Refs.~\cite{gallardo1988,gallardo,Ring_1983}. The values of deformation parameters are taken from the finite-range droplet model (FRDM) data~\cite{FRDM_2012}, which is $\beta_2 = 0$ for both the nuclei.
\begin{table*}
    \centering
    \begin{tabular}{c c} 
    \hline
    \hline
     Basis Label &   Basis Size  \\
    \hline
    Basis 1 & ~$N$ = 0 to 10 \\
    Basis 2 & $N$ = 2 to 8 \\
    Basis 3 & $N$ = 3 to 6 \\
    Basis 4 & $N$ = 4 to 6 \\
    Basis 5 & $N$ = 4 to 5 \\
    \hline
    \hline
    \end{tabular}
    \caption{The label of different basis sizes used in the results shown in Fig.~\ref{fig:peak_width_sn}}
    \label{basis}
\end{table*}

\begin{figure}
     \centering
     \begin{subfigure}[b]{0.51\textwidth}
    \centering
    \includegraphics[width = \textwidth]{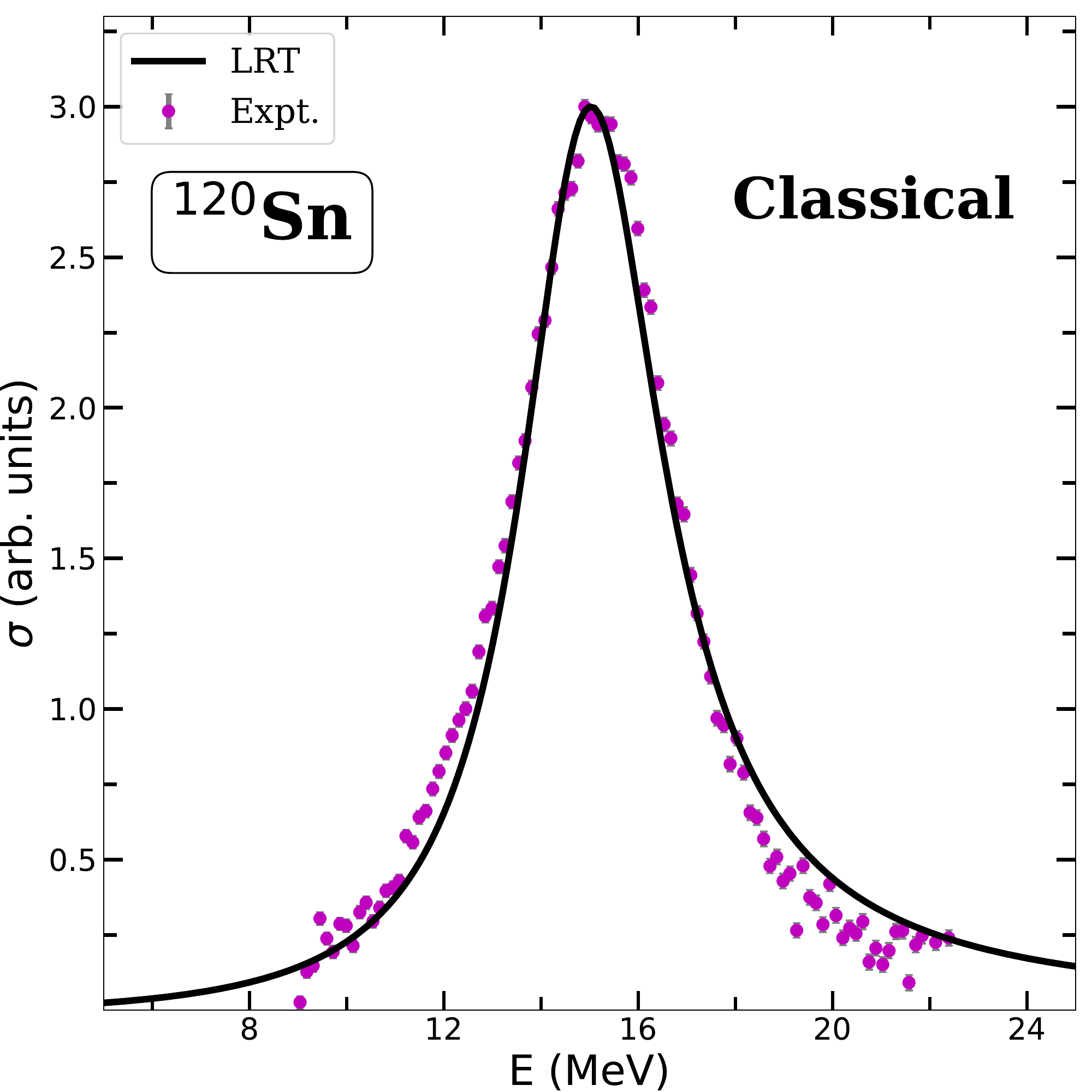}
    \caption{Classical computation. }
    \label{fig:Sn_classical}
    \end{subfigure}
     \hspace{-1.16cm}
     \begin{subfigure}[b]{0.51\textwidth}
     \centering
    \includegraphics[width = \textwidth]{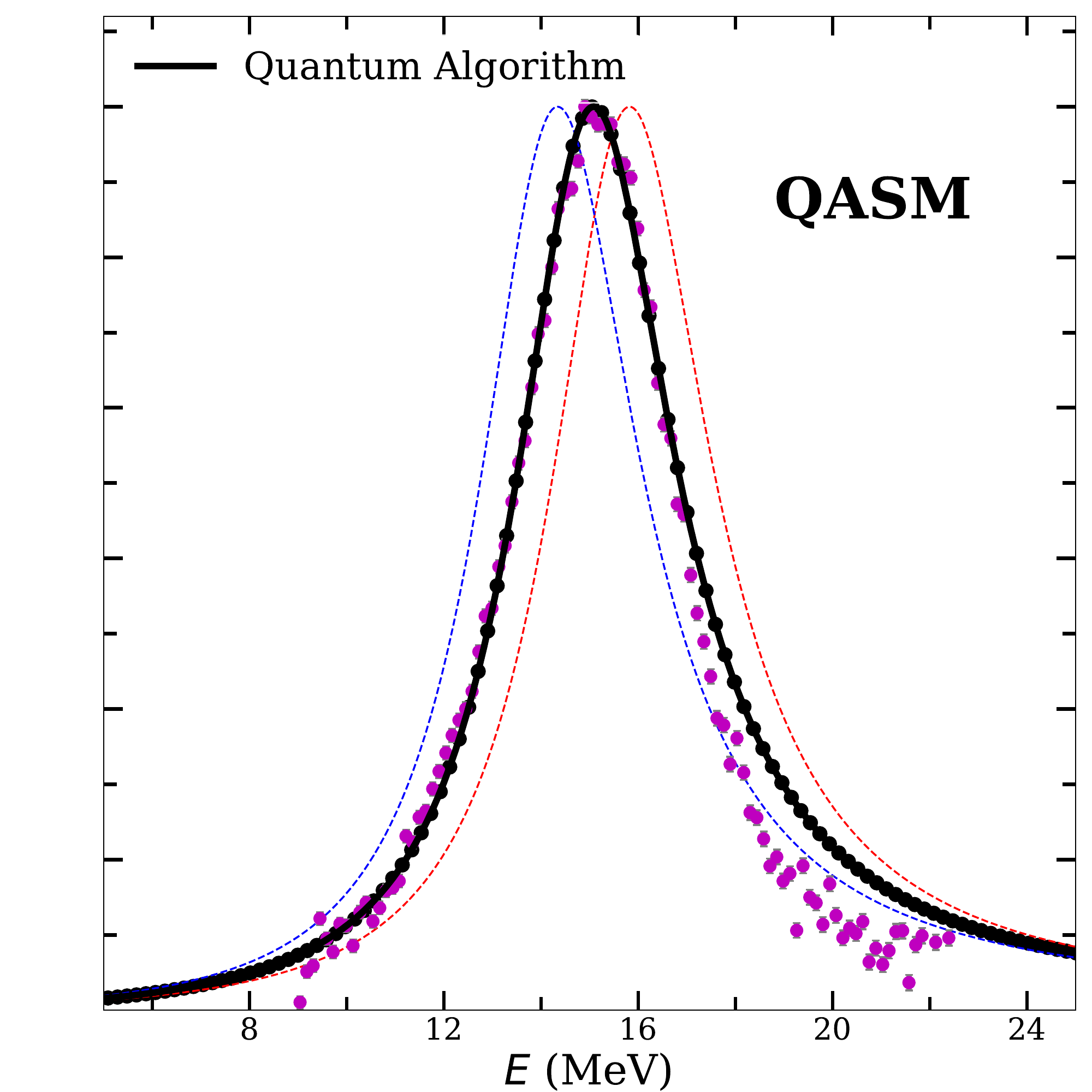}
    \caption{Quantum simulation.}
    \label{fig:Sn_quantum}
    \end{subfigure}
    \caption{For $^{120}$Sn GDR cross-section is compared with the experimental results taken from Ref.~\cite{exp1berman} obtained with: (a) the classical computation within the linear response theory, and (b) the quantum algorithm using the harmonic oscillator (H.O.) Hamiltonian, the red and blue dashed lines show the shift in the GDR peak due to the error in the values of overlaps calculated with the quantum algorithm. }
    \label{fig:Sn}
\end{figure}

In Fig.~\ref{fig:Sn}, we show our results for $^{120}$Sn where Fig~\ref{fig:Sn_classical} shows the results obtained using the classical computation. These results are obtained using the single-particle states [Eq.~(\ref{R0})] and give a good agreement with the experiment. The quantum computing results for $^{120}$Sn are shown in Fig.~\ref{fig:Sn_quantum}.  The results are obtained using the QASM simulator~\cite{Qasm} available in the IBM Qiskit package~\cite{Qiskit}. The QASM simulator simulates the real behavior of a quantum computer. For the QASM results, we use 100 independent runs of our quantum algorithm (Hamiltonian simulation + overlap calculation) with 8000 shots of the simulator for each data point. 

We get a very good agreement with the response obtained on the quantum computer. There is a slight overestimation of GDR response in the energy range of 18 - 20 MeV in the results of QASM. Still, we get a good corroboration with the experimental data. The blue and red dashed lines show the shifted GDR peaks in the QASM results because of the error in the value of the overlaps in the SWAP test. These overlaps represent the probability of transition from the ground state to the excited state, and any change in these values corresponds to an overall shift in the GDR peak ($E_0$). The error in the overlap values is introduced because of the inherent quantum randomness. The shift in the $E_0$ can be corrected by changing the value of $\kappa$ but if we keep the parameter $\kappa$ constant for every run of the quantum algorithm, then the peak shifts in energy, leading to an overall error ($\Delta E_0$) in the GDR peak energy. Due to the sensitivity of the mean and standard deviation to the outliers in the data, we consider the median absolute deviation (MAD) for the measure of errors. The error in the values of energies and wave functions obtained by the LCU approach does not significantly affect the GDR peak in the QASM result. 
The interesting point is that the GDR peak is sensitive to even small errors resulting from the usual randomness of quantum mechanics built into the QASM simulator.  

We are currently limited to only spherical nuclei as for a deformed system, the degeneracy of the basis states is lifted, and we thus require a much larger number of basis states to calculate the response function, which leads to an increase in the number of qubits required to perform the calculations on the quantum computer.

\begin{figure}
     \centering
     \begin{subfigure}[b]{0.51\textwidth}
    \centering
    \includegraphics[width = \textwidth]{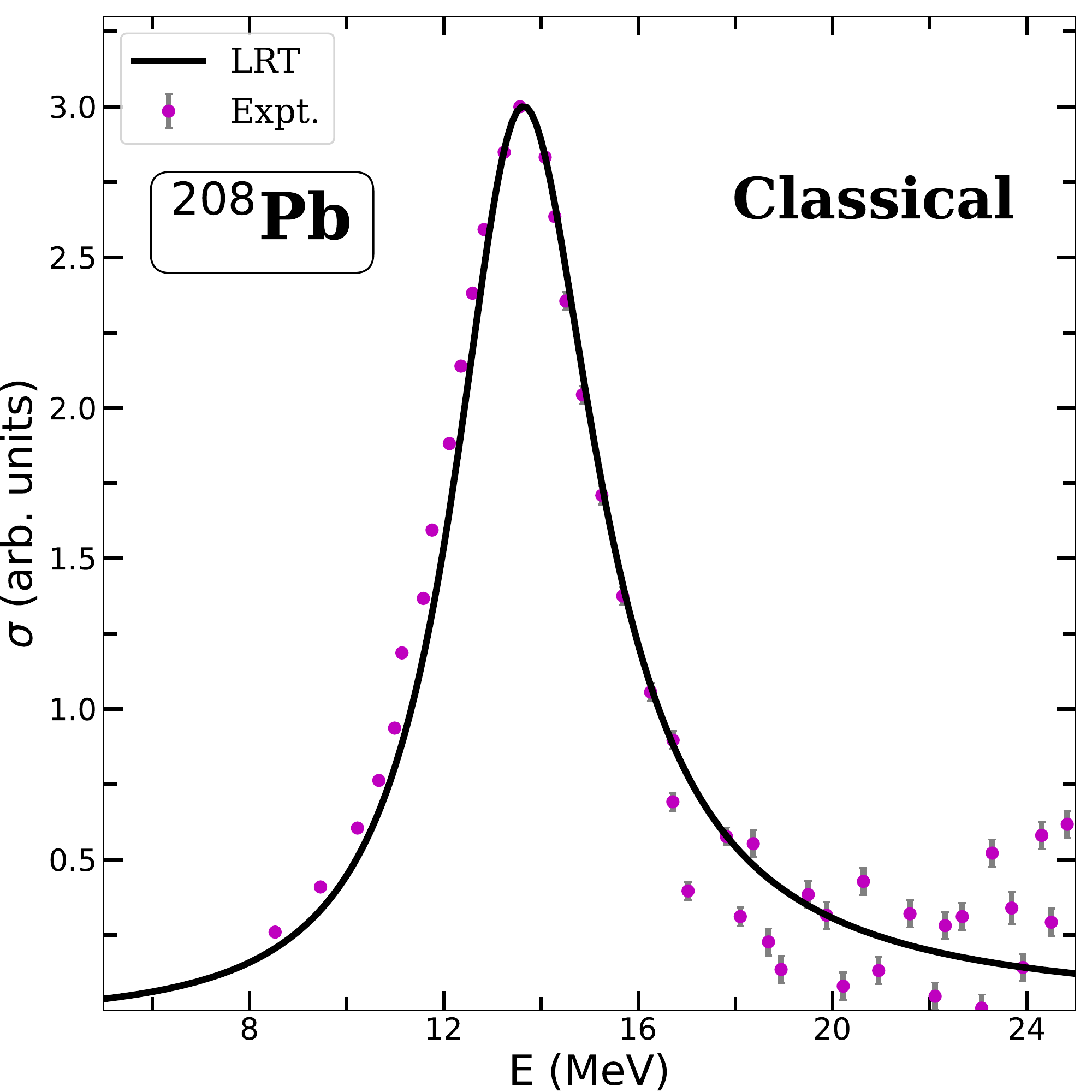}
    \caption{Classical computation. }
    \label{fig:Pb_classical}
    \end{subfigure}
     \hspace{-1.16cm}
     \begin{subfigure}[b]{0.51\textwidth}
     \centering
    \includegraphics[width = \textwidth]{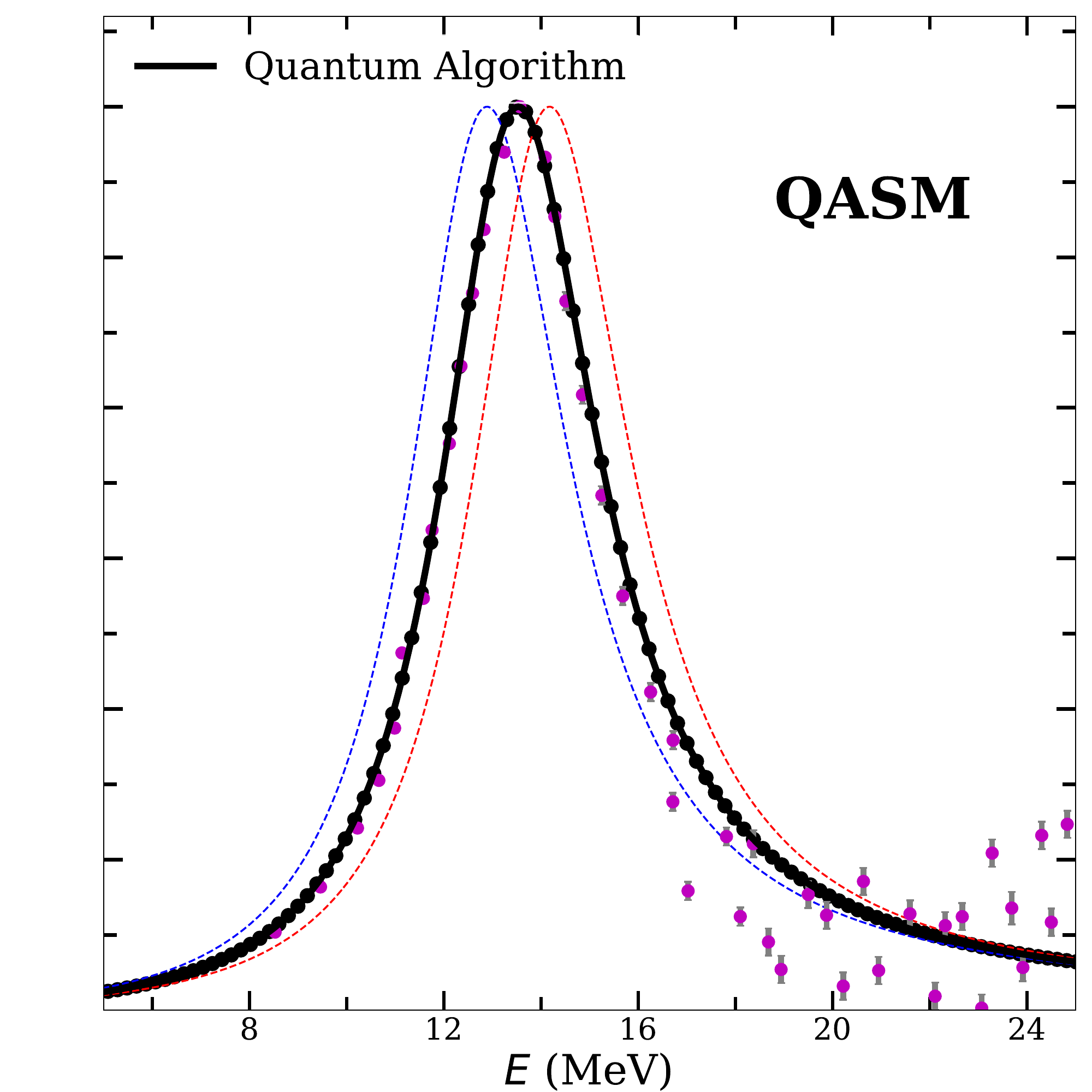}
    \caption{Quantum simulation.}
    \label{fig:Pb_quantum}
    \end{subfigure}
    \caption{Same as Fig.~\ref{fig:Sn} but for $^{208}$Pb. }
    \label{fig:Pb}
\end{figure}

\begin{figure}
    \centering
    \includegraphics[width=0.65\textwidth]{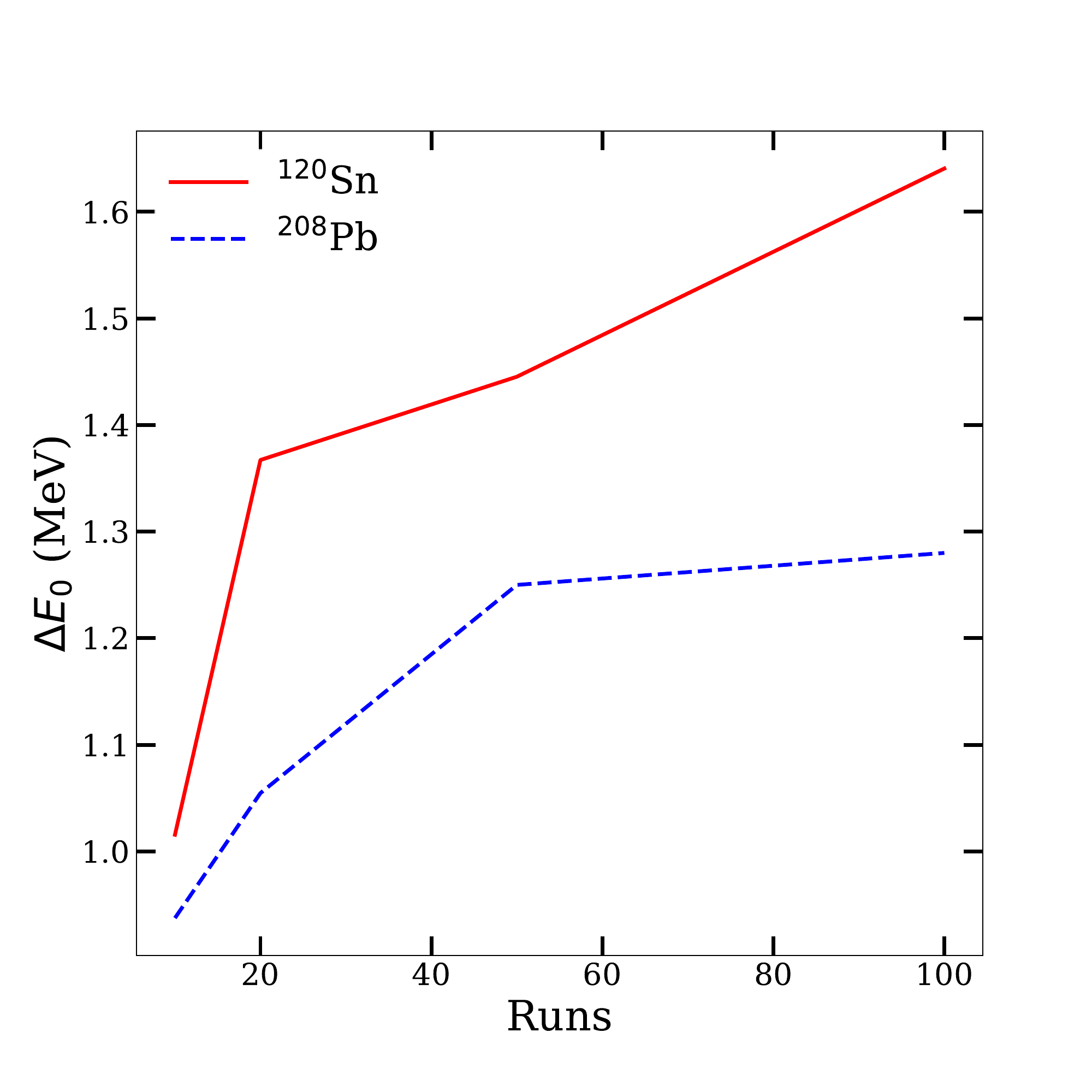}
    \caption{The value of error ($\Delta E_0$) in the GDR peak energy ($E_0$) as a function of the number of runs of the quantum algorithm for both nuclei. This error is introduced due to the quantum randomness in the QASM results. }
    \label{fig:error_run}
\end{figure}

\begin{figure}
    \centering
    \includegraphics[width=0.65\textwidth]{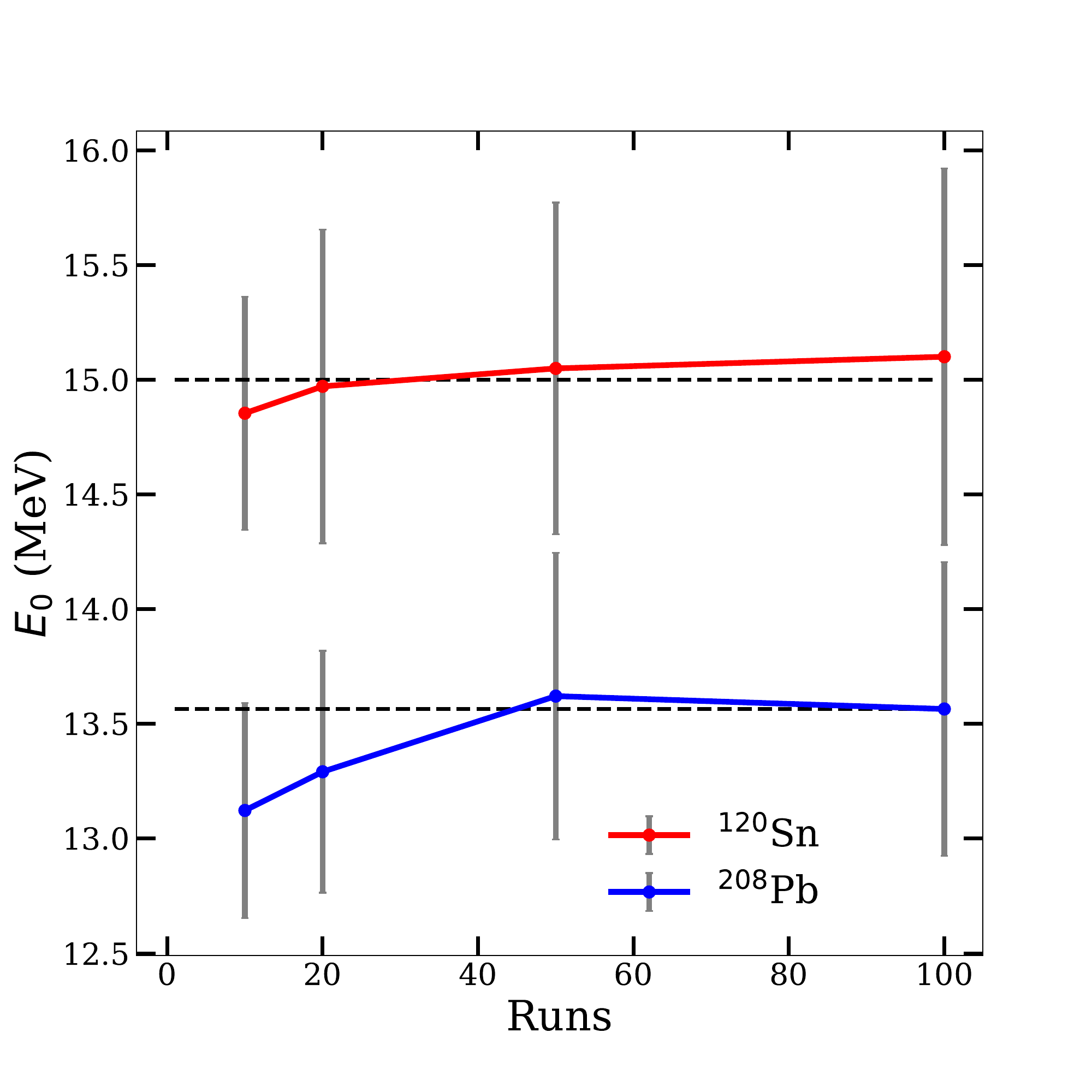}
    \caption{GDR peak energy ($E_0$) as a function of the number of runs of the quantum algorithm. The horizontal black dashed lines show each nucleus's corresponding experimental values of $E_0$.}
    \label{fig:peak_run}
\end{figure}

In Fig.~\ref{fig:Pb}, we show our results for $^{208}$Pb to benchmark our outcomes in the Pb region. Similar to $^{120}$Sn, we get a good agreement with the experiment in the case of classical computation results (Fig.~\ref{fig:Pb_classical}). On the other hand, the QASM results also match the experimental data well. However, similar to $^{120}$Sn, we also see the sensitivity of GDR peak energy to the error in QASM results. In Fig.~\ref{fig:error_run}, we show the error in GDR peak ($\Delta E_0$) as a function of the number of runs used in calculating the GDR cross-section, and we see that this error increases initially with the number of runs and then saturates as more runs mean more information about the inherent randomness. $^{120}$Sn results show a bigger jump in the value of $\Delta E_0$, although the error value remains below the value of 2 MeV for both nuclei and shows saturation as we increase the number of runs.  This can be better understood by looking at the GDR peak energy $E_0$ as a function of runs, as shown in Fig.~\ref{fig:peak_run}. The value of $E_0$ for both nuclei approaches the true experimental value (horizontal dashed lines) as we increase the number of runs, whereas the error in the peak energy saturates.

Overall, our quantum approach for the nuclear response function gives good results, which explains the experiment in both the Sn and Pb regions. Extension of this approach to the deformed nuclei where Nilsson model like Hamiltonians can be utilized to obtain the response can prove a useful tool for the experimental studies in GDR.

\section{CONCLUSIONS}\label{conclude}
We employed an algorithm based on the linear combination of unitaries (LCU) to simulate the Hamiltonian and utilized the SWAP test to obtain the expectation value of the dipole operator (overlap), which is used in computing the nuclear response. We have utilized the QASM simulator to simulate the real quantum computer and compared the results of our simulation with the classical computation of the response obtained within the linear response theory based on random phase approximation.  Our results agree with the experimental data in both the Sn and Pb regions. The GDR peak is sensitive to the even small error resulting from the usual randomness of quantum mechanics that are built into the QASM simulator. We have chosen a smaller basis size which is concentrated around the Fermi level in both nuclei, as we have shown that only states near the Fermi level contribute the most to the nuclear response. Our approach requires more than twice the number of qubits required for representing the nuclear state in qubit space, as the overlap circuit must simultaneously represent the ground and excited states. Further optimization to calculate the dipole matrix elements on a quantum computer, such as a destructive SWAP test~\cite{swap:2013}, can improve the requirement of qubits. 

Extending this approach to a deformed system is straightforward, however, it will increase the number of qubits as deformed states require a larger basis size. Currently, we are limited by the number of qubits for the current era of quantum computers. Still, this approach gives direct access to the excited states of a nuclear system under the influence of small external perturbation and is quite successful in reproducing the experimental results, and can be a very useful tool for interpreting the experimental data for the nuclear response.

\begin{acknowledgments}
This work was supported in part by the U.S.~Department of Energy, Office of Science, Office of High Energy Physics, under Award No.~DE-SC0019465. 
\end{acknowledgments}
% \bibliography{refrences}
%merlin.mbs apsrev4-1.bst 2010-07-25 4.21a (PWD, AO, DPC) hacked
%Control: key (0)
%Control: author (8) initials jnrlst
%Control: editor formatted (1) identically to author
%Control: production of article title (-1) disabled
%Control: page (0) single
%Control: year (1) truncated
%Control: production of eprint (0) enabled
%

\end{document}